\documentclass{article}

\usepackage{arxiv}

\usepackage[utf8]{inputenc} 
\usepackage[T1]{fontenc}    
\usepackage{hyperref}       
\usepackage{url}            
\usepackage{booktabs}       
\usepackage{amsfonts}       
\usepackage{nicefrac}       
\usepackage{microtype}      
\usepackage{lipsum}		
\usepackage{graphicx}
\usepackage{natbib}
\usepackage{doi}
\usepackage{threeparttablex}
\usepackage{sistyle}
\SIthousandsep{,}

\usepackage{import}

\def\bz{{\mathbf z}}
\def\bZ{{\mathbf Z}}
\def\bbeta{{\boldsymbol{\beta}}}
\def\btheta{{\boldsymbol{\theta}}}

\def\blambda{{\boldsymbol{\lambda}}}

\usepackage{amsthm}

\usepackage{booktabs}
\usepackage{amsmath}

\usepackage{mathtools}
 
\usepackage{dsfont}

\def\bZ{\pmb{Z}}
\def\bz{\pmb{z}}
\def\bbeta{\pmb{\beta}}
\def\blambda{\pmb{\lambda}}
\def\I{\textrm{I}}
\def\E{\textrm{E}}
\def\btheta{\pmb{\theta}}
\allowdisplaybreaks

\title{Improved computational efficiency and stability when imputing censored covariates: Analytic and numerical approaches}

\author{ \href{https://orcid.org/0000-0001-5380-2427}{\includegraphics[scale=0.06]{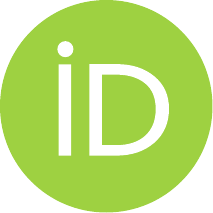}\hspace{1mm}Sarah C.~Lotspeich} \\
	Department of Statistical Sciences\\
	Wake Forest University\\
	Winston-Salem, NC 27109 \\
	\texttt{lotspes@wfu.edu} \\
	\And
	\href{https://orcid.org/0000-0002-6112-9030}{\includegraphics[scale=0.06]{orcid.pdf}\hspace{1mm}Ethan M.~Alt} \\ 
        Department of Biostatistics \\
	University of North Carolina at Chapel Hill \\ 
        Chapel Hill, NC 27599 \\
	\texttt{ethanalt@live.unc.edu} \\
}

\renewcommand{\shorttitle}{Improved computational efficiency and stability when imputing censored covariates}

\hypersetup{
pdftitle={Speedy CMI},
pdfsubject={q-bio.NC, q-bio.QM},
pdfauthor={Sarah C.~Lotspeich, Ethan M.~Alt},
pdfkeywords={accelerated failure-time model, coarsened data, conditional survival function, multiple imputation, parametric survival modeling, proportional hazards model},
}

\begin{document}
\maketitle

\begin{abstract}
Imputation is a popular approach to handling censored, missing, and error-prone covariates -- all coarsened data types for which the true values are unknown. However, there are nuances to imputing these different data types based on the mechanism dominating the unobserved values and other available information. For example, in prospective studies, the time to a disease diagnosis will be incompletely observed if only some patients are diagnosed by the end of the follow-up. Some will be randomly right-censored, and patients' disease-free follow-up times must be incorporated into their imputed values.  Assuming noninformative censoring, censored values are replaced with their conditional means, which are calculated by estimating the conditional survival function of the censored covariate and then integrating over it. Semiparametric approaches are common, which estimate the survival with a Cox model and then the integral with the trapezoidal rule. While these approaches offer robustness, they come at the cost of computational efficiency and stability in numerically approximating an improper integral. After modeling the survival function parametrically, we derive analytic solutions for conditional mean imputed values under many common distributions. We define stabilized calculations for other distributions. Parametric imputation using various distributions and calculations is implemented in the R package, \textit{speedyCMI}.
\end{abstract}


\section{Motivation}
\label{sec:motivation}
\subsection{Censored Covariates}
At first glance, a censored covariate may look no different than a missing covariate or one with measurement error. After all, censored, missing, and error-prone covariates all belong to the family of \textit{coarsened} data types, and -- in any case -- we do not know the true covariate value. However, a censored covariate has (i) partial information about the true value that cannot be ignored and (ii) a censoring mechanism that can impact estimation and inference. Thus, nuance is required to handle censored, rather than missing or error-prone, covariates. Still, there are common repercussions of ignoring these problems in our data, and the statistical strategies to overcome them overlap substantially. 

Suppose we are interested in fitting a normal linear regression model between an outcome and a censored covariate. If we ignore the censoring, we could ``plug in'' the observed values and fit the usual model. However, this ``naive'' model will lead to biased parameter estimates, inflated variability, and uncontrolled type-I error rates in hypothesis testing \cite{AustinBrunner2003,AustinHoch2004}. A straightforward alternative is the complete case analysis, which subsets to only the observations with uncensored covariates. However, it is inefficient and can be inconsistent unless the covariate censoring is independent of the outcome \cite{AshnerGarcia2023}. Therefore, more advanced methods are needed to obtain efficient, unbiased statistical inference with a censored covariate. 

\subsection{Existing Methods} \label{subsec:exmethods}

Design-based approaches, like inverse probability weighting (IPW), can offer consistency in more situations than the complete case analysis with similar computational ease \cite{MatsouakaAtem2020}. However, as censored observations are still deleted in the analysis, IPW will be similarly inefficient. Augmented IPW can offer improved statistical efficiency, but its adoption for censored covariates is far behind that for missing and error-prone ones. These design-based methods are robust, as they just require estimating the weights, which can be done with nonparametric or semiparametric approaches. 

There are also model-based approaches to handling censored covariates. In general, they offer greater statistical efficiency than design-based approaches by incorporating more information from the censored observations. Under proper specification,  parametric maximum likelihood estimators (MLEs) \cite{Atem&Matsouka2017} offer the greatest statistical efficiency and other desirable asymptotic properties, including normality. Semiparametric MLEs offer better statistical efficiency than design-based estimators and more robustness than their parametric counterparts \cite{Gomezetal2003}. 
Still, implementing MLEs is taxing because each new setting requires a new algorithm to be derived.

Imputation, whereby censored covariates are replaced by an informative placeholder, offers similar benefits to maximum likelihood estimation (e.g., Atem, Qian, Maye, Johnson, and Betensky 2017; Atem, Sampene, and Greene 2019). This method is approachable to statistical or nonstatistical audiences, as imputation is a common solution to missing data. When covariates are censored, they are commonly imputed with their conditional means given the censored value and any additional fully observed covariates. Under standard assumptions, calculating these conditional means requires (i) estimating the conditional distribution of the censored covariate and (ii) integrating over the corresponding survival function. The survival function is frequently estimated semiparametrically (e.g., with a Cox model) and its integral approximated numerically (e.g., with the trapezoidal rule).

When imputing censored covariates, there are also unique challenges because the area under the survival curve is of interest. Thus, while these semiparametric imputation approaches offer robustness, it comes at the cost of computational efficiency and stability, since the integrals are not analytically tractable and require numerical approximations. Extrapolating the baseline hazard beyond the last observed event is nontrivial and requires unverifiable assumptions. Recently, Lotspeich and Garcia (2025) demonstrated that semiparametric imputation approaches could suffer from high bias, due to the trapezoidal rule only being defined to integrate up to the last observed event. 

\subsection{Considerations With Parametric Imputation Approaches}

The choice between parametric and semiparametric modeling generally comes down to favoring efficiency versus robustness, respectively. When imputing censored covariates, however, the robustness in modeling its distribution semiparametrically can be overpowered by the computational difficulties and potential bias in numerically approximating the integral. Herein, we focus on promoting the usability of parametric imputation for censored covariates through the development of novel analytic and numerical approaches, offering improved computational efficiency and stability over existing semiparametric ones. We also empirically validate a simple imputation model selection procedure. 

With our computational improvements, there are advantages to a well-specified parametric imputation approach. It may be preferable in settings where semiparametric imputation fails, for example, under extra heavy censoring when more extrapolation of the tail of the survival curve might be needed. Also, having accessible parametric comparators is key to the development of nonparametric and semiparametric methods. 

\subsection{Overview}

We propose new analytic and numerical approaches to parametric conditional mean imputation (CMI) of censored covariates that offer improved computational stability and efficiency, sometimes with better statistical precision. Under many common distributions, we derive the analytic solutions for the conditional mean imputed values. We also define stabilized integral calculations for distributions without analytically tractable solutions. Through simulations, we quantify the sizable computational gains of our analytic and stabilized solutions over the existing conditional mean formula seen in the literature. We also empirically validate a simple and effective imputation model selection procedure. Motivated by the Framingham teaching dataset, we demonstrate how to select an imputation model and apply the proposed methods, which are implemented in the \textit{speedyCMI} R package.

\section{Methods}
\label{sec:methods}

\subsection{Model and Data}\label{subsec:model&data}
Consider a continuous outcome $Y$, censored continuous covariate $X$, and other fully observed covariates $\bZ$ which are assumed to be related through the normal linear regression model
$Y = \beta_0 + \beta_1 X + \bbeta_{2}^T\bZ + \epsilon,$ where errors $\epsilon$ are assumed to follow a mean-zero normal distribution with variance $\sigma^2$. Primary interest is in statistical inference of the parameters $\bbeta = (\beta_0, \beta_1, \bbeta_{2}^{\rm T}, \sigma^2)^{\rm T}$, but, unfortunately, $X$ is censored and thus incompletely observed. 

We focus on right, random censoring in $X$, but modifications needed for other censoring types can be found in Section~\ref{sec:generalize}. In this setting, we observe the value $W = \min(X, C)$ in place of $X$, with event indicator $\Delta = \I(X \leq C)$. Thus, observations for the $n$ subjects in the sample are captured as \{($Y_i, \Delta_i, W_i, \bZ_i$), $i \in \{1, \dots, n\}$\}. 

Under the assumption of \textit{noninformative censoring}, $C$ and $X$ are conditionally independent given additional covariates $\pmb{Z}$. In practice, if $X$ is noninformatively right censored by $C$, then $C$ only dictates the lower bound of possible $X$ values. In probability statements, this assumption is expressed as ${\rm P}(X, X>C|\pmb{Z})={\rm P}(X>C|\pmb{Z})$. The assumption of noninformative censoring simplifies the calculation of the conditional mean imputed values for censored covariates. 

\subsection{Conditional Mean Imputation} \label{sec:imp}
In conventional missing data settings, imputation is a popular approach to obtain valid statistical inference without sacrificing the power of the full sample. Thus, it is a promising method to handle censored covariates with the exception that censored data are not fully missing. For a censored subject $i$ ($i \in \{1, \dots, n\}$), the covariate $X_i$ may be unobserved but it is known to be greater than the observed value $W_i$ under right censoring. This partial information about $X_i$ (namely, that $X_i > W_i$) distinguishes a censored covariate from a missing or error-prone one and can be captured through a conditional mean imputation (CMI) approach \cite{Richardson&Ciampi2003}.

For ease of exposition, we assume that the covariate is right censored, but it is straightforward to generalize our approach to left- and interval-censored covariates. With CMI, a randomly right-censored covariate value $W_{i}$ is replaced with its conditional mean: 
\begin{align}
\E(X|X>W_i,\bZ_i) &= W_i + \frac{\int_{W_i}^{\infty}S(x|\bZ_i)dx}{S(W_i|\bZ_i)}, \label{exp_right}
\end{align}
where $S(x|\bz) = \Pr(X > x|\bZ=\bz)$ is the conditional survival function for $X$ given $\bZ = \bz$ evaluated at $x$. For a randomly right censored covariate, \eqref{exp_right} was first introduced in Atem, Qian, Maye, Johnson, and Betensky (2017), and Lotspeich, Grosser, and Garcia (2022) includes a thorough derivation. Conveniently, \eqref{exp_right} can be recognized as the sum of the observed covariate value $W_i$ and the mean residual life (MRL) at that value given the subject's additional covariates $\bZ = \bZ_i$. Notably, this form hinges upon the noninformative censoring assumption. Additional derivations for the conditional means under left or interval censoring are in the Supplementary Materials (Section~S.1). 

Calculating the conditional means in \eqref{exp_right} is done in two steps: estimating the conditional survival function of the censored covariate and then integrating over it from the censored value to infinity. Existing approaches to each of these steps are summarized below. 
\begin{itemize}
    \item[]\textit{Step 1 (Survival):} Thus far, nonparametric and semiparametric approaches, like the Kaplan--Meier estimator and Cox proportional hazards model, respectively, have been the predominant choice to estimate $S(x|\bz)$ \cite{AtemEtAl2017,AtemSampeneGreene2019,AtemEtAl2016}. These approaches avoid distributional assumptions about $X$ given $\bZ$, but this robustness can come at a cost. Step functions like the Kaplan--Meier estimator and Breslow's estimator (used to estimate the survival function from a Cox model of the hazard) are defined on the observed data. Therefore, integrating over their tail up to infinity in Step 2 can be computationally intensive and require some additional assumptions to do well (e.g., extrapolation of the survival function beyond the last observed event). Finally, the semiparametric Cox model approaches still require the proportional hazards assumption to hold, which becomes less tenable as the covariate space increases, particularly when additional covariates $\bZ$ are continuous. 
    \item[]\textit{Step 2 (Integral):} After estimating $S(x|\bz)$ with these popular approaches, the trapezoidal rule is commonly used to approximate the integral from the smallest to largest observed values, $W_{(1)}$ to $W_{(n)}$, respectively \cite{AtemEtAl2017,AtemSampeneGreene2019,AtemEtAl2016}. The trapezoidal rule is computationally simple, and this approach can be relatively accurate for the integral in \eqref{exp_right} when the last observed $W_{(n)}$ is uncensored. However, when $W_{(n)}$ is censored, this trapezoidal rule approximation will lead to biased estimates of the conditional means, as the area under $S(x|\bz)$ for $x > W_{(n)}$ will not be included \cite{LotspeichGarcia2025}. 
\end{itemize}

After replacing all of the censored covariate values $W_i$ with the corresponding conditional means from Steps 1--2, we can employ the usual regression modeling techniques on the ``complete'' (i.e., imputed) dataset for inference about $\bbeta$. We focus on continuous outcomes, so the CMI estimators $\hat{\pmb{\beta}}$ are found through ordinary least-squares regression. The asymptotic consistency of $\hat{\pmb{\beta}}$ in linear regression has been proven under proper specification of the analysis and imputation models \cite{Bernhardtetal2015}. 

\subsection{Estimation of the Conditional Survival Function}\label{efficient_survival}

While non- and semiparametric CMI approaches for a randomly right-censored covariate have been more common in recent years \cite{AtemEtAl2016,AtemEtAl2017,AtemSampeneGreene2019}, at least one parametric version has been proposed. In it, Royston (2007) assumed that the censored covariate followed a normal distribution, which is limited in its applications. 

We instead construct a parametric CMI approach that accommodates a number of plausible distributions for the censored covariate, broadening its applications. An estimation step is still needed to calculate the conditional means, since the parameters $\btheta$ governing the survival function $S_{\btheta}(x|\bz)$ are likely unknown in practice but can be obtained using an approach such as maximum likelihood estimation. For practical use, we recommend using existing software to obtain $\hat{\btheta}$ whenever possible. See the Supplementary Materials (Section~S.2) for a list of select software choices. \phantom{\cite{Royston2007}}

\subsection{Analytic Solutions for Common Distributions}\label{analytic_solutions}

For some popular distributions, analytic solutions exist for the conditional mean in \eqref{exp_right}, offering higher computational efficiency and better numerical stability than numerical integration. We focus on the common setting of a right-censored covariate, relegating  derivations and formulas under left or interval censoring to the Supplementary Materials. 

\subsubsection{Weibull distribution}\label{weibull}
In regression, the Weibull distribution is a convenient choice because it has a closed-form survival function and also because it belongs to the accelerated failure time and proportional hazards families. Assuming $X$ given $\bZ_i$ has a Weibull distribution with shape $\alpha > 0$ and scale $\lambda_i > 0$ (where $\lambda_i$ is a function of $\bZ_i$), the conditional mean imputed value is defined: 
\begin{align}
    \E(X|X>W_i,\bZ_i) &= W_i + \frac{\Gamma(1/\alpha) S_{\Gamma}(\lambda_{i} W_i^{\alpha} | 1/\alpha)}{\exp\left(-\lambda_{i} W_i^{\alpha}\right)\alpha \lambda_{i}^{1/\alpha}},
    \label{eq:truncmean_weibull_no_int}
\end{align}
where $\Gamma(a, t) = \int_{t}^{\infty} x^{a - 1} \exp(-x) dx$ is the (upper) incomplete Gamma function and $S_{\Gamma}(t | a)$ is the survival function at $t$ of a gamma random variable with shape $a > 0$ and scale equal to $1$. The upper incomplete Gamma and survival functions needed to compute \eqref{eq:truncmean_weibull_no_int} are implemented in standard statistical packages (e.g., R's built-in \texttt{pgamma} function). The derivation of \eqref{eq:truncmean_weibull_no_int} and analytic formulas for the conditional means under left or interval covariate censoring can be found in the Supplementary Materials (Section~S.3.1). 

\subsubsection{Exponential distribution}\label{expo}

The exponential distribution, another popular choice, is a special case of the Weibull distribution. The shape parameter is fixed at $\alpha = 1$, providing simplified expressions for key quantities (e.g., the survival function). When this assumption is reasonable, substituting $\alpha = 1$ into \eqref{eq:truncmean_weibull_no_int}, the conditional mean formula reduces to the conditional mean under an exponential distribution: $\E(X|X>W_i,\bZ_i)  = W_i + 1/\lambda_i$. However, the assumption that $\alpha = 1$ implies a constant hazard $h(x) = \lambda_i$ over time $x$, which may not be realistic. 

\subsubsection{Log-normal distribution}
The Weibull distribution's primary limitation is its monotone hazard function. Some random variables require a hump-shaped hazard (i.e., it initially increases, reaches a maximum, then decreases). In these situations, the log-normal distribution is a good option. Assuming $X$ given $\bZ_i$ has a log-normal distribution with location $\mu_i > 0$ (where $\mu_i$ is a function of $\bZ_i$) and scale $\sigma > 0$, the conditional mean imputed value is defined $\E(X |X > W_i, \bZ_i)$
\begin{align}
    = \exp\left(\mu_i + \frac{\sigma^2}{2}\right) \left[\frac{1 - \Phi\left\{\frac{\log(W_i) - \mu_i}{\sigma} - \sigma \right\} }{1 - \Phi\left\{ \frac{\log (W_i) - \mu_i}{\sigma} \right\} }\right],
    \label{eq:lognormal_expectation}
\end{align}
which is easily evaluated using functions for the normal cumulative distribution function (CDF) $\Phi(\cdot)$ in standard statistical software (e.g., R's built-in \texttt{pnorm} function). The derivation of \eqref{eq:lognormal_expectation} and conditional means under left or interval covariate censoring can be found in the Supplementary Materials (Section~S.3.2). 

\subsubsection{Log-logistic distribution}\label{loglogistic}
The log-logistic distribution is another common choice in survival regression, due to its flexible hazard function, which can be decreasing or hump-shaped. It also approximates the log-normal distribution well while having fatter tails; this connection is appealing because the log-normal does not have a closed-form hazard or survival function, while the log-logistic does. Assuming $X$ given $\bZ_i$ has a log-logistic distribution with shape $\alpha > 1$ and scale $\lambda_i > 0$ (where $\lambda_i$ is a function of $\bZ_i$), the conditional mean imputed value is $\E\left(X |X > W_{i}, \bZ_i\right) $
\begin{align}
&= W_i + 
    \frac{\lambda_i}{\alpha} \left\{ 1 + \left( \frac{W_i}{\lambda_i} \right)^{\alpha} \right\} B\left(\frac{\alpha - 1}{\alpha}, \frac{1}{\alpha}\right) F_{\beta}\left(\left\{ 1 + \left( \frac{W_i}{\lambda_i} \right)^{\alpha} \right\}^{-1} \Bigg| \frac{\alpha - 1}{\alpha}, \frac{1}{\alpha}\right), \label{final_condl_mean_loglog}
\end{align}
where $B(a, b) = \int_{0}^{1} x^{a-1} (1 - x)^{b-1}dx$ is the beta function with parameters $a$ and $b$ and $F_{\beta}(t | a, b)$ is the CDF at $t$ of a beta random variable with shapes $a > 0$ and $b > 0$. All terms in \eqref{final_condl_mean_loglog} can be computed using built-in functions of R, like \texttt{pbeta}.

For shapes $\alpha \leq 1$, the conditional mean assuming a log-logistic distribution does not exist, i.e., $\E\left(X |X > W_{i}, \bZ_i\right) \to \infty$, presenting a practical limitation. Since $\alpha > 1$ corresponds to a hump-shaped hazard, the log-logistic distribution is not useful for CMI if the hazard is monotone (i.e., strictly increasing or decreasing). In such a case, the Weibull distribution might be preferred. The Supplementary Materials contains the derivation of \eqref{final_condl_mean_loglog} and analytic formulas for the conditional means under left or interval censoring (Section~S.3.3). 

\subsubsection{Piecewise exponential distribution}
\label{sec:pwe}
A lesser-used but quite powerful distribution for survival models is the piecewise exponential (PWE). Rather than assuming that the hazard is constant across all $x$ values, like the exponential distribution (Section~\ref{expo}), the PWE distribution assumes that the hazard is only constant within small sub-intervals (i.e., the hazard is \emph{piecewise} constant). When using many sub-intervals, the PWE distribution can approximate any baseline hazard. However, this flexibility comes at a cost; namely, the hazard is still assumed to be constant beyond the last sub-interval. Although, in principle, other assumptions could be incorporated.

Suppose that we discretize the time axis for $X$ into $J$ disjoint sub-intervals, say, $[\tau_{j-1}, \tau_j)$ for $j \in \{1, \dots, J\}$ and $0 = \tau_0 < \cdots < \tau_J = \infty$. The interval boundaries $\tau_j$ are fixed, commonly chosen as percentiles. Assuming a PWE distribution with rates $\blambda_i = (\lambda_{i1}, \dots, \lambda_{iJ})$ (where all $\lambda_{ij}$ are functions of $\bZ_i$), the conditional mean imputed value is $ \E\left( X | X > W_i, \bZ_i \right)$
\begin{align}
  & = W_i + \frac{1}{ S(W _i|\bZ_i)}\left\{\frac{ S(W_i|\bZ_i) - S(\tau_{J_i}|\bZ_i) }{\lambda_{iJ_i}} + 
    \sum_{j = J_i + 1}^J \frac{ S(\tau_{j-1}|\bZ_i) - S(\tau_j|\bZ_i) }{\lambda_{ij}}\right\},\label{cond_mean_pwe}
\end{align}
where $J_i$ denotes the index of the interval into which $W_i$ falls and $\lambda_{ij}$ denotes the constant hazard for interval $[\tau_{j-1}, \tau_j)$ given covariates $\bZ_i$, i.e., $\lambda_{ji} = h(x|\bZ_i)$ for $\tau_{j-1} \leq x < \tau_j$. 
Derivations for \eqref{cond_mean_pwe} along with analytic solutions for left and interval are provided in the Supplementary Materials (Section~S.3.4). 

\subsection{Stabilized Integral Calculations for Other Distributions}\label{stabilize_integral}

There are many distributions for which the key integral $\int_{W_{i}}^{\infty}S(x|\bZ_i) dx$ in the conditional mean formula will not be available in terms of known quantities. In these situations, numerical integration can be used to estimate in the conditional means. However, approximations of integrals with infinite bounds can be quite unstable; we propose ways to stabilize them. 

If the mean $\E(X|\bZ_i)$ is known or can be well approximated (e.g., via Monte Carlo techniques), we can utilize the following identity to stabilize the integral calculation:
\begin{align}
    \int_{W_i}^{\infty} S(x|\bZ_i) dx = \int_{0}^{\infty} S(x|\bZ_i) dx - \int_{0}^{W_i} S(x|\bZ_i) dx = \E\left(X|\bZ_i\right) - \int_{0}^{W_i} S(x|\bZ_i) dx, \label{int_mean_approx}
\end{align}
which reduces to numerically evaluating an integral that is absolutely continuous over the compact interval $[0, W_i]$ rather than $[W_i, \infty)$. Thus, plugging \eqref{int_mean_approx} into \eqref{exp_right} yields our first stabilized formula for the conditional mean: 
\begin{align}
\E(X|X>W_i,\bZ_i) &= W_i + \frac{1}{S(W_i|\bZ_i)}\left\{\E\left(X|\bZ_i\right) - \int_{0}^{W_i} S(x|\bZ_i) dx\right\}. \label{exp_right_alt}
\end{align}

When $\E(X|\bZ_i)$ is unknown and cannot be well approximated, we instead stabilize the integral via the following transformation:
\begin{align}
  \int_{W_i}^{\infty} S(x|\bZ_i) dx &= \int_{0}^{1} \exp\left(- \left[H\left\{W_i + (1 - t)/t|\bZ_i\right\} + 2 \log (t)\right] \right) dt,
  \label{eq:numericalintegral}
\end{align}
where $H(t|\bz) = \int_{0}^{t} h(t|\bz) dt$ is the conditional cumulative hazard function for $X$ given $\bZ = z$ evaluated at $t$. Note that as $t \to 0$, $H\{W_i + (1 - t) / t|\bZ_i\} \to \infty$, while $2 \log (t) \to -\infty$. Hence, this transformation ``balances'' the integrand to avoid over/underflow in the tails, and the conditional means can be estimated by plugging \eqref{eq:numericalintegral} into \eqref{exp_right} for our second stabilized formula for the conditional mean. The integral in \eqref{eq:numericalintegral} can also be evaluated on the log-scale for even more stability (Section~S.4 in the Supplementary Materials).

\subsection{Generalizing to an Interval Censored Covariate}\label{sec:generalize}
More generally, we can impute an \emph{interval-censored} covariate, where we know $L_i < X \leq U_i$ for some lower bound $L_i$ and upper bound $U_i$. As detailed in the Supplementary Materials (Section S.1), a covariate censored on $(L_i, U_i]$ can be replaced with $\E(X|L_{i} < X \leq U_{i}, \bZ_i)$
\begin{align}
    &= \frac{\int_{L_{i}}^{U_{i}} x f(x | \bZ_i) dx }{S(L_{i} | \bZ_i) - S(U_{i} | \bZ_i)}
    = \frac{L_{i} S(L_{i} | \bZ_i) - U_{i} S(U_{i} | \mathbf{Z}_i) + \int_{L_{i}}^{U_{i}} S(x | \bZ_i) dx}{S(L_{i} | \bZ_i) - S(U_{i} | \mathbf{Z}_i)},
    \label{eq:intervalcens_gen}
\end{align}
where the second equality is obtained via integration by parts. If $L_i = W_i$ and $U_i = \infty$ (i.e., right censoring), \eqref{eq:intervalcens_gen} reduces to \eqref{exp_right} provided $\lim_{x\to\infty} x S(x) = 0$, which is typical.

\subsection{Multiple Imputation Framework}\label{sec:multimp}

When singly imputing with conditional means particular attention has to be given to the standard error (SE) estimator. As noted in Richardson and Ciampi (2003) and others, the usual SE estimators (e.g., those accompanying an ordinary least-squares regression model) can underestimate the true variability of the CMI estimators. To ensure that the uncertainty around the conditional mean imputed values is captured appropriately, multiply imputing with conditional means may be preferred. Most often, bootstrapping is employed \cite{AtemSampeneGreene2019}, but resampling the imputation model parameters has also been proposed \cite{LotspeichGrosserGarcia2022}. Final analysis model estimates and SEs are found using Rubin's rules to pool estimates from the multiple imputations. 

\section{Simulation Studies}
\label{sec:sims}
To examine how the different conditional mean calculations performed in terms of computational speed and operating characteristics, we conducted simulations. R code for the simulations and parametric CMI approach can be found in the \textit{speedyCMI} package, available on GitHub at \url{https://github.com/sarahlotspeich/speedyCMI}. 

\subsection{Setup and Data Generation}

Samples of $n \in \{500, 1000, 2500\}$ subjects were generated in the following way. First, a fully observed binary covariate $Z$ was generated from a Bernoulli distribution with $\Pr(Z = 1) = 0.5$. From this, the true covariate $X$ was generated from a log-normal distribution with location $\mu= 0.05Z$ and scale $\sigma= 0.5$ (on the log-scale). The continuous outcome was generated as $Y = 1 + 0.5X + 0.25Z + \epsilon$, with $\epsilon$ independently generated from a standard normal distribution. To randomly right censor $X$, $C$ was generated from an exponential distribution with rates $q \in \{0.2, 0.7\}$, corresponding to light ($\approx 20\%$) and heavy ($\approx 50\%$) censoring rates, respectively. Finally, the observed covariate value $W = \min(X, C)$  and event indicator $\Delta = \I(X \leq C)$ were constructed.  

We considered parametric CMI where the conditional means were calculated using (i) the \textit{original integral} approximation from \eqref{exp_right}, (ii) the \textit{stabilized integral (with the mean)} from \eqref{exp_right_alt}, (iii) the \textit{stabilized integral (without the mean)} from \eqref{eq:numericalintegral}, and (iv) the \textit{analytic solution} from \eqref{eq:lognormal_expectation}. As a comparator, semiparametric CMI was included, which was implemented in the \textit{imputeCensRd} package \cite{LotspeichGrosserGarcia2022}. Each setting was replicated 1000 times, fitting the linear regression outcome model to data imputed using each of the five methods (four parametric and one semiparametric CMI approaches). Results from all replications were summarized together with respect to the following. 

The per-replicate computing times on a 2020 Macbook Pro (M1 Chip) with 16 gigabytes of memory were measured using the \texttt{"elapsed"} time returned from the \texttt{system.time} function in base-R. They were summarized per-setting by the mean and median (for robustness). The total computing times across all replicates were also reported. The empirical bias and percent bias for $\beta_1$ (the parameter on censored covariate $X$) was calculated for each setting as the mean of the per-replicate biases (i.e., the differences between the estimated and true parameters). The empirical SEs for $\beta_1$ were also included, alongside the average SE estimators (i.e., the empirical mean of the estimated SEs) and empirical coverage probability (CP) for 95\% confidence intervals (CIs). The empirical relative efficiency for $\beta_1$ was calculated as the ratio of the variances of the per-replicate estimated parameters from the CMI approaches to the full cohort analysis (i.e., with no censored values). Larger relative efficiency indicated greater amounts of information recovered through imputation. 
 
\subsection{Single Imputation Simulations}\label{sims:single_imp}

\begin{figure}[ht]
    \caption{Average computing runtime per replicate for single imputation simulations. Solid and dashed lines connect the mean and median per-replicate computing times, respectively.
\label{figure:single-imputation-average-runtime}}
    \centering 
    \includegraphics[width=\textwidth]{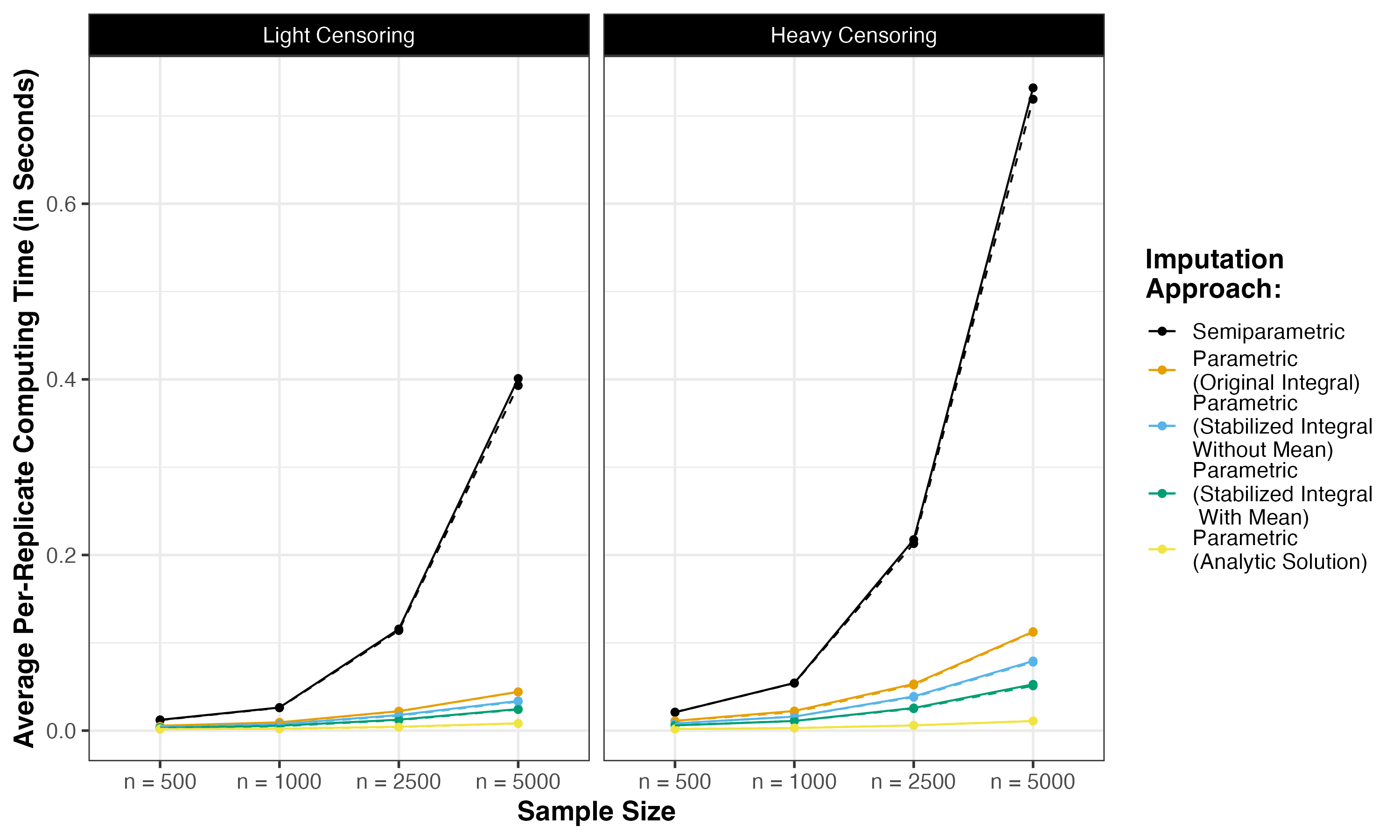}
\end{figure}

For all sample sizes and censoring rates considered, semiparametric CMI had the slowest computing times, followed by parametric CMI using the original integral (Figure~\ref{figure:single-imputation-average-runtime}). Of the three proposed parametric CMI calculations, the stabilized integral (without the mean) offered some computational reprieve, but the most noticeable gains in computing speed were seen for the stabilized integral (with the mean) and the analytic solution. 

The relative computational ease of parametric CMI  using the analytic solution was more advantageous with larger sample sizes and under heavier censoring. In general, this approach scaled up remarkably well, continuing to offer similarly low runtimes of approximately $0.01$ second or less, on average, regardless of the demands of the data (e.g., the sample size or censoring rate). These observations all held when looking instead at the \textit{total} computing runtime across \num{1000} replicates (Supplemental Figure~S1). 

Estimates for $\beta_1$ using any of the four parametric CMI approaches were identical (Supplemental Figure~S2). As expected, parametric CMI was empirically unbiased for $\beta_1$, and became more statistically efficient as either (i) sample size increased (for fixed censoring rate) or (ii) censoring rate decreased (for fixed sample size) (Supplemental Table~S1). Somewhat surprisingly, the naive SE estimator after single imputation performed well here, closely approximating the empirical SE and leading to CPs close to the nominal $95\%$. Semiparametric CMI was similarly unbiased but could be slightly less efficient (by up to 5\%).

\subsection{Multiple Imputation Simulations}\label{subsec:mi-sims}

For the multiple imputation simulations, only the more computationally intensive setting with heavy censoring was considered, and the sample size was fixed at $n=1000$ subjects. Now, the number of imputations was varied with $B \in \{1, 5, 10, 20, 40\}$, where $B = 1$ is the single imputation setting from Section~\ref{sims:single_imp}.

\begin{figure}[ht]
    \caption{Average computing runtime per-replicate for simulations with an increasing number of imputations $B$, assuming heavy censoring and $n = 1000$ subjects. Solid and dashed lines connect the mean and median per-replicate computing times, respectively. 
\label{figure:multiple-imputation-average-runtime}}
    \centering 
    \includegraphics[width=\textwidth]{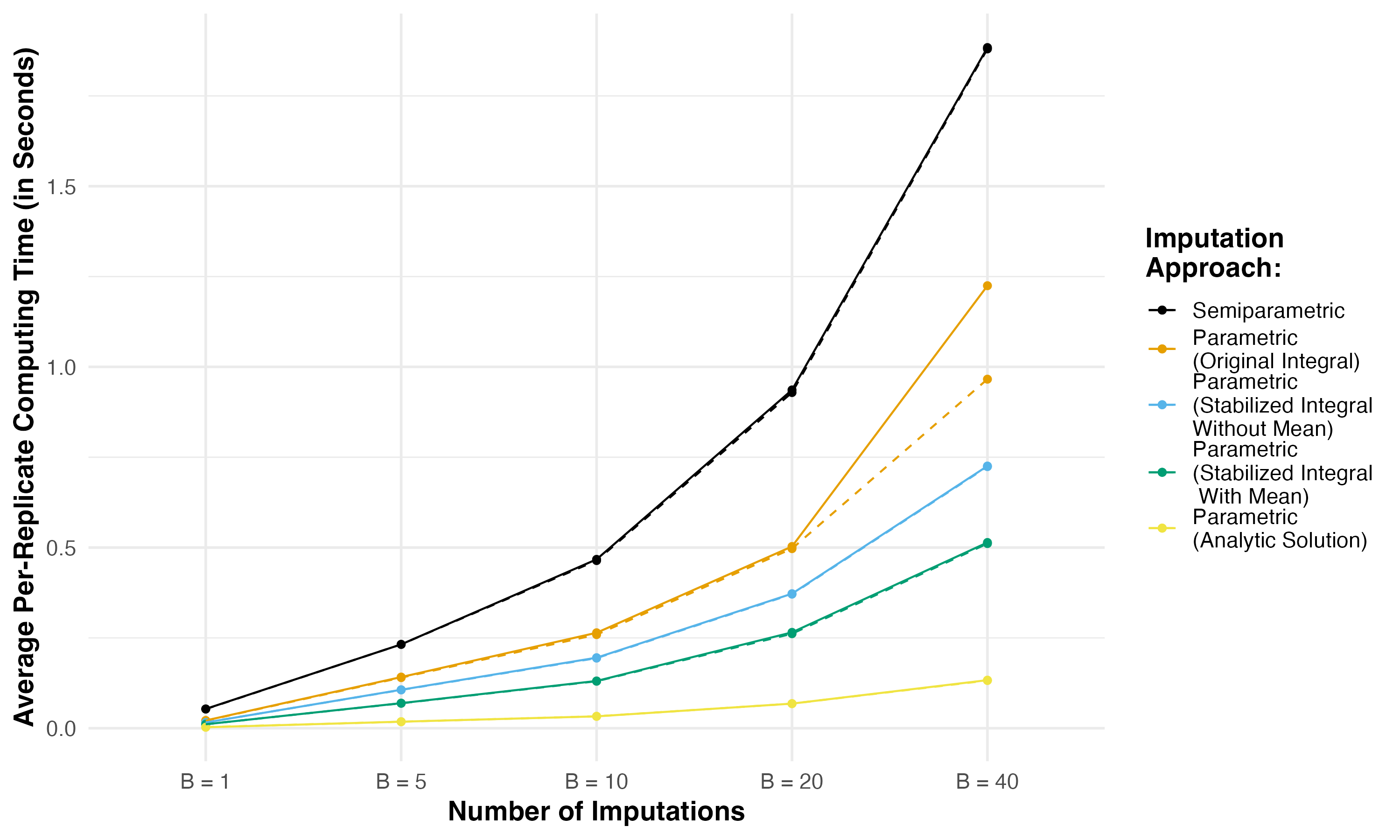}
\end{figure}

The rankings from computationally slowest to fastest were the same in single or multiple imputation and stable as $B$ increased (Figure~\ref{figure:multiple-imputation-average-runtime}). Still, the gaps between approaches widened as the number of imputations $B$ increased. In particular, parametric CMI with the analytic solution scaled up extremely well, with average per-replicate computing times $\leq 0.1$ second for as many as $B = 40$ imputations. With the largest number of imputations, the other approaches took between $0.5$ and $1.9$ seconds per replicate, on average. The graph of total computing runtimes tells the same story (Supplemental Figure~S3). 

Estimates for $\beta_1$ were similar to those under single imputation and essentially unchanged as the number of imputations $B$ increased (Supplemental Figure~S4). However, the pooled SE estimator for multiple imputation (parametric or semiparametric) using Rubin's rules was slightly larger, on average, than the empirical (Supplemental Table~S2), leading to conservative $95\%$ CIs  (CPs $98-99\%$). Since we fit an analysis model for $Y$ given $X$ and $Z$ but impute from $X$ given $X>C$ and $Z$ (rather than $X$ given $Y$ and $Z$ as with a missing covariate), noncongeniality could explain high CPs \cite{Meng1994}. While single imputation had fine SE estimates before, we include multiple imputation because it is commonly employed with semiparametric CMI (e.g., Atem, Matsouaka, and Zimmern 2019). 

\subsection{Model Misspecification Simulations}\label{subsec:sims-misspec}

Proper specification is key when adopting parametric approaches like this one. For the data generating mechanism of log-normal $X$ used thus far, samples of $n = 1000$ subjects were simulated under heavy censoring. In these data, we explored the impact of misspecifying the imputation model as (i) log-logistic, (ii) Weibull, (iii) exponential, or (iv) PWE with $J = 10$ disjoint sub-intervals based on the percentiles of the uncensored $X$ values. For each distribution, we considered both single imputation and multiple imputation ($B = 10$). 

As expected, parametric CMI assuming the log-normal distribution (the correct specification) led to the lowest bias in estimating $\beta_1$ (Table~\ref{table:misspec}). Still, the PWE distribution performed nearly as well as the true one. The log-logistic and Weibull distributions were less than $5\%$ and $8\%$ biased, respectively. The exponential distribution was glaringly biased ($47\%$), which is not surprising since it makes the strictest assumptions. In comparison, bias of parametric CMI assuming the log-normal or PWE distributions was better than semiparametric CMI ($2.8\%$), while the log-logistic and Weibull were slightly worse. For the four distributions that were reasonably unbiased (excluding exponential), the SE estimators performed similarly, with multiple imputation still leading to conservative estimates. 

\begin{table}
\caption{Estimates of $\beta_1$, the parameter on the censored covariate $X$, from the full cohort (i.e., all uncensored) and conditional mean imputation (CMI) analyses assuming different distributions for the imputation model.
\label{table:misspec}}
\centering
\resizebox{\columnwidth}{!}{
\begin{threeparttable}
\begin{tabular}{llrrclrrcccclrrcccc}
\hline 
& & \multicolumn{3}{c}{\textbf{Full Cohort}} && \multicolumn{6}{c}{\textbf{Parametric CMI}} && \multicolumn{6}{c}{\textbf{Semiparametric CMI}} \\
\cmidrule(l{3pt}r{3pt}){3-5} \cmidrule(l{3pt}r{3pt}){7-12} \cmidrule(l{3pt}r{3pt}){14-19}
\textbf{Distribution} && \textbf{Bias} & \textbf{(\%)} & \textbf{SE} && \textbf{Bias} & \textbf{(\%)} & \textbf{ESE} & \textbf{ASE} & \textbf{CP}& \textbf{RE} && 
\textbf{Bias} & \textbf{(\%)} &  \textbf{ESE} & \textbf{ASE} & \textbf{CP}&  \textbf{RE}\\ \hline 
\addlinespace
\multicolumn{19}{c}{\textbf{Multiple Imputation $\pmb{(B = 10)}$}} \\
\addlinespace
Exponential &  & $0.002$ & ($0.34$) & $0.052$ &  & $-0.235$ & ($-47.02$) & $0.048$ & $0.064$ & $0.018$ & $1.166$ &  & $0.014$ & ($2.80$) & $0.083$ & $0.109$ & $0.988$ & $0.394$\\
Weibull &  & $0.002$ & ($0.34$) & $0.052$ &  & $ 0.038$ & ($  7.58$) & $0.085$ & $0.114$ & $0.986$ & $0.370$ &  & $0.014$ & ($2.80$) & $0.083$ & $0.109$ & $0.988$ & $0.394$\\
Log-Normal \textit{(Truth)} &  & $0.002$ & ($0.34$) & $0.052$ &  & $ 0.004$ & ($  0.82$) & $0.080$ & $0.106$ & $0.984$ & $0.422$ &  & $0.014$ & ($2.80$) & $0.083$ & $0.109$ & $0.988$ & $0.394$\\
Log-Logistic &  & $0.002$ & ($0.34$) & $0.052$ &  & $-0.024$ & ($ -4.76$) & $0.076$ & $0.101$ & $0.974$ & $0.461$ &  & $0.014$ & ($2.80$) & $0.083$ & $0.109$ & $0.988$ & $0.394$\\
Piecewise Exponential &  & $0.002$ & ($0.34$) & $0.052$ &  & $ 0.005$ & ($  1.07$) & $0.081$ & $0.107$ & $0.988$ & $0.410$ &  & $0.014$ & ($2.80$) & $0.083$ & $0.109$ & $0.988$ & $0.394$\\
\addlinespace
\multicolumn{19}{c}{\textbf{Single Imputation}} \\
\addlinespace
Exponential &  & $0.002$ & ($0.34$) & $0.052$ &  & $-0.234$ & ($-46.84$) & $0.046$ & $0.044$ & $0.001$ & $1.267$ &  & $0.012$ & ($2.35$) & $0.079$ & $0.075$ & $0.936$ & $0.432$\\
Weibull &  & $0.002$ & ($0.34$) & $0.052$ &  & $ 0.038$ & ($  7.52$) & $0.081$ & $0.079$ & $0.914$ & $0.405$ &  & $0.012$ & ($2.35$) & $0.079$ & $0.075$ & $0.936$ & $0.432$\\
Log-Normal \textit{(Truth)} &  & $0.002$ & ($0.34$) & $0.052$ &  & $ 0.004$ & ($  0.80$) & $0.076$ & $0.074$ & $0.942$ & $0.462$ &  & $0.012$ & ($2.35$) & $0.079$ & $0.075$ & $0.936$ & $0.432$\\
Log-Logistic &  & $0.002$ & ($0.34$) & $0.052$ &  & $-0.024$ & ($ -4.85$) & $0.073$ & $0.070$ & $0.928$ & $0.504$ &  & $0.012$ & ($2.35$) & $0.079$ & $0.075$ & $0.936$ & $0.432$\\
Piecewise Exponential &  & $0.002$ & ($0.34$) & $0.052$ &  & $ 0.006$ & ($  1.11$) & $0.077$ & $0.074$ & $0.940$ & $0.449$ &  & $0.012$ & ($2.35$) & $0.079$ & $0.075$ & $0.936$ & $0.432$\\
\bottomrule
\end{tabular}
\begin{tablenotes}[flushleft]
\item{\em Note:} \textbf{Bias (\%)}: empirical bias (empirical percent bias); \textbf{ESE}: empirical standard error; \textbf{ASE}: average standard error estimator; \textbf{CP}: empirical coverage probability of 95\% confidence intervals; \textbf{RE}: empirical relative efficiency to the full cohort analysis. 
All entries are based on \num{1000} replicates.
\end{tablenotes}
\end{threeparttable}
}
\end{table}

\subsection{Model Selection Simulations}\label{subsec:sims-select}

We used simulations from Section~\ref{subsec:sims-misspec} to demonstrate a potential imputation model selection procedure. Let $k$ denote the number of parameters estimated for a candidate model and $\mathcal{L}$ denote the maximum of that model's likelihood. The Akaike information criterion, $AIC = 2k - 2\log(\mathcal{L})$, and Bayesian information criterion, $BIC = k\log(n) - 2\log(\mathcal{L})$, were calculated for each candidate model. Smaller $AIC$ and $BIC$ values were preferred. 

\begin{figure}[ht]
    \caption{Model diagnostics Akaike information criterion (AIC) and Bayesian information criterion (BIC) for the imputation model of $X$ given $Z$ when the truth was log-normal. 
\label{figure:diagnostics}}
    \centering 
    \includegraphics[width=\textwidth]{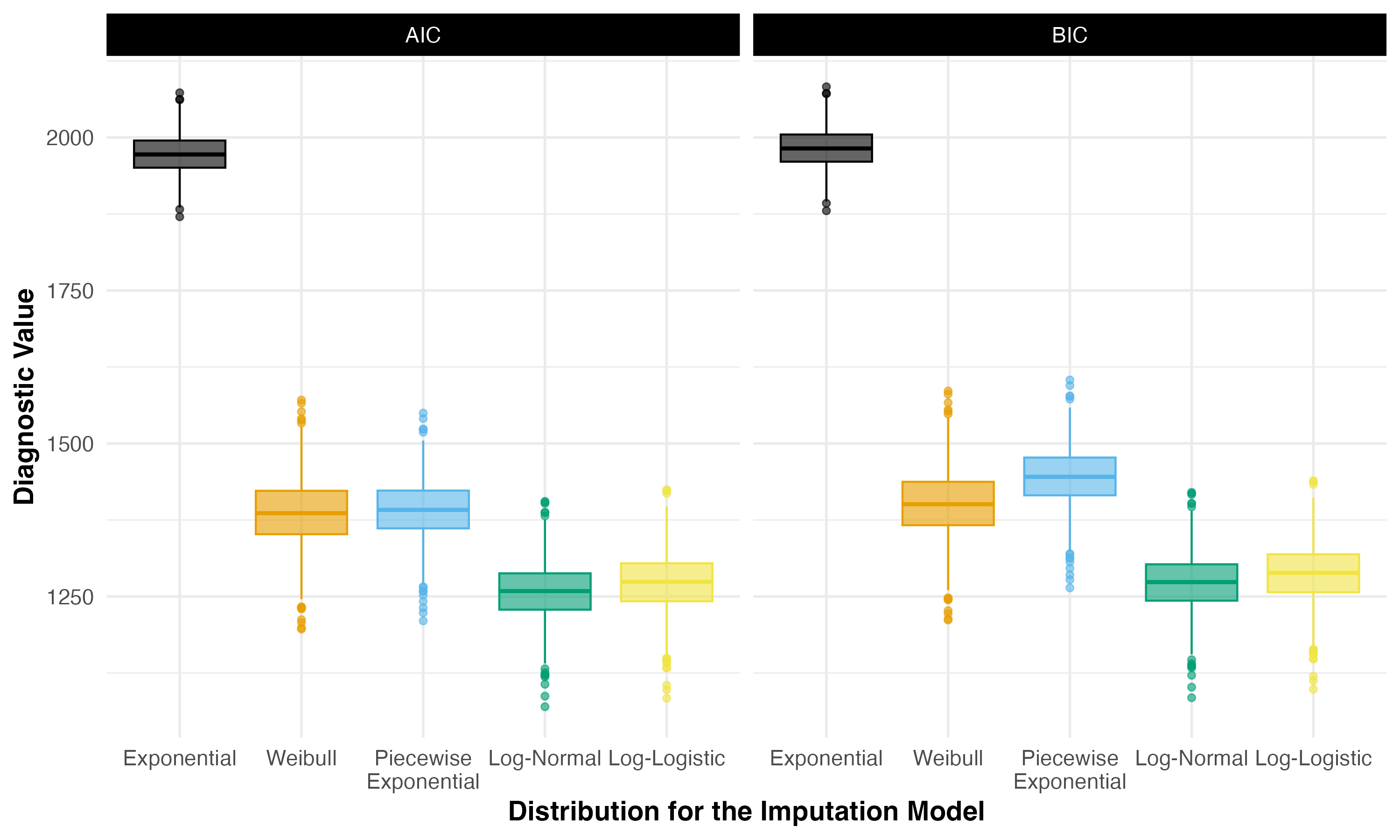}
\end{figure}

The distributions of $AIC$ and $BIC$ across the \num{1000} replicates were similar for nearly all candidate distributions (Figure~\ref{figure:diagnostics}). As expected, $AIC$ and $BIC$ for the log-normal distribution (the correct specification) were smallest, on average. The log-logistic was not far behind, which was reasonable since it approximates the log-normal. According to either diagnostic, the worst fit was the exponential distribution; this result was also expected since the true data generation process has a non-constant baseline hazard. 
Only the PWE distribution ranked differently, on average, based on $AIC$ versus $BIC$, which was attributed to its added complexity, since it had the most parameters (i.e., largest $k$) and thus was penalized by $BIC$ more than the other distributions. 

For a given dataset, suppose we fit all five candidate imputation models and then used the one with the lowest $AIC$ or $BIC$ for the final analysis. Across the \num{1000} replicates, this procedure  would correctly select the log-normal distribution in $982$ ($98\%$). In the remaining replicates, it would select the log-logistic distribution instead, which -- while not exactly correct -- is closely related and tended to have very similar diagnostics (Figure~\ref{figure:diagnostics}). 

\section{Application to the Framingham Teaching Dataset}\label{sec:framingham}

The Framingham Study is a landmark study in epidemiology, which prospectively studies cardiovascular diseases and their risk factors. It has previously been used to investigate the relationship between the ages of cardiovascular disease onset for participants and their parents using the ``Offspring Study'' (e.g., Atem, Qian, Maye, Johnson, and Betensky
2017). The Offspring Study data are not publicly available, so we used the Framingham teaching dataset \cite{riskCommunicator} to demonstrate our methods with a related model instead. 

The Framingham teaching dataset contains laboratory, clinical, questionnaire, and cardiovascular event data on a subset of \num{4434} patients from the original study. Patients were excluded from our analysis if they (i) had prevalent hypertension at first visit, (ii) had only one study visit, or (iii) were missing body mass index (BMI) at first visit. These criteria left $n = 2218$ patients, and $1211$ ($55\%$) were diagnosed with hypertension during follow-up. 

Suppose we were interested in the relationship between a patient's systolic blood pressure (SBP) at their first visit ($SBP_0$, in millimeters of mercury [mmHg]) and their SBP at either hypertension diagnosis or last visit ($SBP_1$, in mmHg). When nearing a hypertension diagnosis, patients may begin new medical or lifestyle treatments. Thus, a key predictor for $SBP_1$ is the time to hypertensive status ($TIME$). Moreover, $TIME$ is correlated with $SBP_0$ and $SBP_1$, so failing to adjust for it, a confounder, could lead to downstream bias.

We further adjusted for sex ($FEMALE$), $AGE$ (in years), and $BMI$ at first visit. Thus, the model of interest was a linear regression: 
\begin{align}
    SBP_1 &= \beta_0 + \beta_1 SBP_0 + \beta_2 TIME + \beta_3 FEMALE + \beta_4 AGE + \beta_5 BMI + \epsilon,
    \label{eq:mod_framingham}
\end{align}
where $\beta_1$ described how BP changes, on average, after controlling for proximity to hypertension diagnosis and sex, age, and BMI at first visit. The $TIME$ to hypertension covariate was randomly right-censored for $1007$ undiagnosed patients ($45\%$). To model SBP change in \eqref{eq:mod_framingham}, $TIME$ to hypertension for patients who had not yet been diagnosed was imputed. 

\subsection{Selecting the Imputation Model}\label{sec:fram-select}

Censored $TIME$ to hypertension diagnosis was imputed with its conditional mean, 
\begin{align}
\E(TIME|TIME>FOLLOWUP, SBP_0,FEMALE,AGE,BMI),  \label{condl_mean_framingham}
\end{align}
where $FOLLOWUP$ denotes disease-free follow-up time in the Framingham teaching dataset. Patients were followed for a maximum of $25$ years, but they could also be censored before that time due to death or other causes. We assumed that $FOLLOWUP$ and $TIME$ were conditionally independent given other covariates such that $TIME$ was noninformatively censored and \eqref{condl_mean_framingham} can be computed following the definition in \eqref{exp_right}. To select a conditional distribution for $TIME$ given the other covariates, we considered each of the distributions from Sections~\ref{analytic_solutions}. For the PWE, $J = 10$ disjoint sub-intervals were defined based on the percentiles of the uncensored $TIME$ values, balancing model flexibility with what the data could support. However, $J$ could also be chosen based on $AIC$ or $BIC$. 

As in Section~\ref{subsec:sims-select}, two model selection criteria, $AIC$ and $BIC$, were used to compare these different distributions, and smaller values were preferred. Among all the models considered, the log-normal fit best according to $AIC$ and $BIC$, while the exponential fit worst (Supplemental Table~S3). Aside from the exponential, the  models had similar mean and median predictions for time to hypertension. Since the proportional hazards models (exponential, Weibull, and PWE) fit the worst, this assumption may not hold here.

The best-fitting imputation model -- the log-normal -- led to median $TIME$ to hypertension diagnosis (after imputing) of $21.1$ years from first visit. Among undiagnosed patients, the distribution was relatively symmetrical around the median of $42.5$ years to hypertension (Supplemental Figure~S5). As expected, the different distributions led to notably different distributions of time to hypertension among undiagnosed patients. Still, the log-normal and log-logistic were very similar, along with the Weibull distribution, to a lesser extent. The exponential and PWE led to much flatter (i.e., less heavy-tailed) distributions. 

\subsection{Results from the Blood Pressure Models}

The single and multiple (with $B = 10$ imputations) parametric CMI analyses assuming a log-normal imputation model were compared to four other approaches: (i) the naive analysis, which uses $TIME$ for uncensored patients and $FOLLOWUP$ for censored ones, (ii) the complete case analysis, which uses $TIME$ for uncensored patients and deletes censored ones, (iii) single semiparametric CMI, and (iv) multiple semiparametric CMI (with $B = 10$). The naive analysis is always expected to be invalid, since it ignores censoring. The complete case analysis can yield consistent, albeit inefficient, estimates when the censoring is independent of the outcome \cite{AshnerGarcia2023}. However, $TIME$ to hypertension diagnosis is likely correlated with $SBP_1$, so the complete case is expected to be biased here. The CMI approaches alternatively assumed that $TIME$ was conditionally independent of $FOLLOWUP$ (given covariates), which seemed more reasonable here. 

\begin{figure}[ht]
    \caption{Coefficient estimates (95\% confidence intervals) from the model of systolic blood pressure based on naive, complete case, and conditional mean imputation (CMI) analyses.
\label{figure:forest}}
    \centering 
    \includegraphics[width=\textwidth]{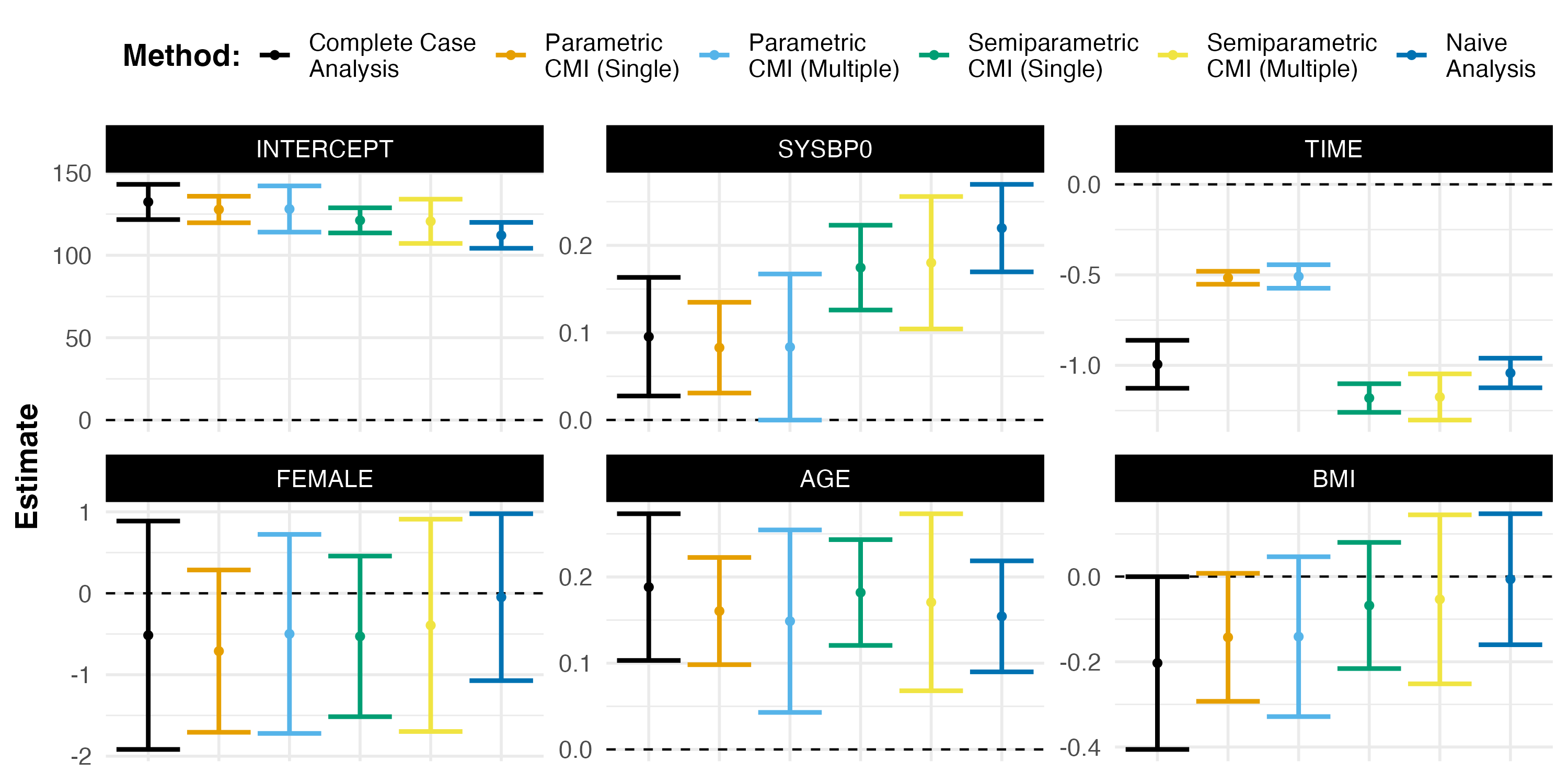}
\end{figure}

We begin by noting findings on which all six estimated models were in agreement. All coefficient estimates shared the same sign (Figure~\ref{figure:forest}). After controlling for all other covariates, there were not residual associations between being $FEMALE$ or $BMI$ at first visit and the outcome, $SBP_1$, but there \emph{were} residual associations between $SBP_1$ and $TIME$ to hypertension diagnosis and $AGE$ at first visit.

For $TIME$, all coefficient estimates were negative and 95\% CIs excluded the null value of zero. These estimates suggested that time to hypertension is an important predictor, as expected, with patients who were farther from a hypertension diagnosis (i.e., with larger $TIME$) expected to have lower SBP, on average, than those who are nearing diagnosis (i.e., with smaller $TIME$). For nearly all methods, $SBP_0$ had a statistically significant positive effect, intuitively suggesting higher baseline BPs yielded higher follow-up BPs, on average, after controlling for other factors. However, for some coefficients (e.g., $SBP_0$ and $TIME$), parametric CMI estimates were smaller in magnitude than those for other analyses. 

Now, while the coefficient estimates using single versus multiple CMI were nearly identical, the approaches differed in the width of their CIs. As seen in simulations (Sections~\ref{subsec:mi-sims}--\ref{subsec:sims-misspec}), the SE estimator for multiple imputation was larger than that for single, leading to wider CIs. In some cases, multiple imputation could lead us to draw different conclusions. 

\section{Discussion}\label{sec:discuss}

As with missing covariates, imputation is an appealing strategy to handle censored ones. While the semiparametric approaches common across the literature can offer robustness, properly specified parametric imputation approaches can offer better computational speed and stability, sometimes with more statistical efficiency. Herein, we derived analytic solutions for the conditional mean imputed values for censored covariates under five common distributions and computationally simpler ``stabilized" calculations for the rest. The computational gains and improved scalability of our proposed approaches were demonstrated through extensive simulation studies and our suggested model selection procedure was illustrated both in simulations and an application to the Framingham teaching dataset. The accompanying \textit{speedyCMI} R package includes implementations for the parametric CMI approaches, along with R scripts for the simulations and Framingham analysis. 

Proper specification is important to fully parametric approaches, although misspecification did not bias the analyses in our simulations much (Section~\ref{subsec:sims-misspec}). If sufficient information is available for us to confidently specify the imputation model for $X$ given $\bZ$, perhaps with a model selection procedure like in Sections~\ref{subsec:sims-select} and \ref{sec:fram-select}, parametric CMI is a very appealing approach to censored covariates. In fact, well-formed  parametric CMI should even rival the efficiency of optimal MLEs with easier adaptability to different model specifications. 

There are many interesting future directions. First, the structured parametric survival curves make this CMI approach a promising way to recover statistical efficiency when $X$ is censored by a limit of detection, while nonparametric and semiparametric ones may not be. Second, our analytic solutions could be used in a Bayesian setting to construct an efficient Gibbs sampler to iteratively impute covariates within the estimation procedure, providing a cohesive framework to reflect uncertainty about parameters and censored covariates.

\bibliographystyle{unsrtnat}
\bibliography{ref}

\begin{thebibliography}{16}
\providecommand{\natexlab}[1]{#1}
\providecommand{\url}[1]{\texttt{#1}}
\expandafter\ifx\csname urlstyle\endcsname\relax
  \providecommand{\doi}[1]{doi: #1}\else
  \providecommand{\doi}{doi: \begingroup \urlstyle{rm}\Url}\fi

\bibitem[Austin and Brunner(2003)]{AustinBrunner2003}
P.~C. Austin and L.~J. Brunner.
\newblock Type {I} error inflation in the presence of a ceiling effect.
\newblock \emph{The American Statistician}, 57\penalty0 (2):\penalty0 97--104,
  2003.

\bibitem[Austin and Hoch(2004)]{AustinHoch2004}
P.~C. Austin and J.~S. Hoch.
\newblock Estimating linear regression models in the presence of a censored
  independent variable.
\newblock \emph{Statistics in Medicine}, 23:\penalty0 411--429, 2004.

\bibitem[Ashner and Garcia(2023)]{AshnerGarcia2023}
M.~C. Ashner and T.~P. Garcia.
\newblock Understanding the implications of a complete case analysis for
  regression models with a right-censored covariate.
\newblock \emph{The American Statistician}, 78\penalty0 (3):\penalty0 335--344,
  2023.

\bibitem[Matsouaka and Atem(2020)]{MatsouakaAtem2020}
R.~A. Matsouaka and F.~D. Atem.
\newblock Regression with a right-censored predictor using inverse probability
  weighting methods.
\newblock \emph{Statistics in Medicine}, 39\penalty0 (27):\penalty0
  4001–4015, 2020.

\bibitem[Atem and Matsouaka(2017)]{Atem&Matsouka2017}
Folefac~D. Atem and Roland~A. Matsouaka.
\newblock Linear regression model with a randomly censored predictor:
  Estimation procedures.
\newblock \emph{Biostatistics and Biometrics Open Access Journal}, 1\penalty0
  (1):\penalty0 1020--1032, 2017.

\bibitem[G{\'o}mez et~al.(2003)G{\'o}mez, Espinal, and Lagakos]{Gomezetal2003}
G.~G{\'o}mez, A.~Espinal, and S.W. Lagakos.
\newblock Inference for a linear regression model with an interval-censored
  covariate.
\newblock \emph{Statistics in Medicine}, 22\penalty0 (3):\penalty0 409--425,
  2003.

\bibitem[Richardson and Ciampi(2003)]{Richardson&Ciampi2003}
David~B. Richardson and Antonio Ciampi.
\newblock Effects of exposure measurement error when an exposure variable is
  constrained by a lower limit.
\newblock \emph{American Journal of Epidemiology}, 157:\penalty0 355--363,
  2003.

\bibitem[Atem et~al.(2017)Atem, Qian, Maye, Johnson, and
  Betensky]{AtemEtAl2017}
F.~D. Atem, J.~Qian, J.~E. Maye, K.~A. Johnson, and R.~A. Betensky.
\newblock Linear regression with a randomly censored covariate: Application to
  an {A}lzheimer's study.
\newblock \emph{Journal of the Royal Statistical Society. Series C (Applied
  Statistics)}, 66\penalty0 (2):\penalty0 313--328, 2017.

\bibitem[Atem et~al.({2019})Atem, Sampene, and Greene]{AtemSampeneGreene2019}
F.~D. Atem, E.~Sampene, and T.~J. Greene.
\newblock Improved conditional imputation for linear regression with a randomly
  censored predictor.
\newblock \emph{Statistical Methods in Medical Research}, 28\penalty0
  (2):\penalty0 432--444, {2019}.

\bibitem[Atem et~al.(2016)Atem, Qian, Maye, Johnson, and
  Betensky]{AtemEtAl2016}
F.~D. Atem, J.~Qian, J.~E. Maye, K.~A. Johnson, and R.~A. Betensky.
\newblock Multiple imputation of a randomly censored covariate improves
  logistic regression analysis.
\newblock \emph{Journal of Applied Statistics}, 43\penalty0 (15):\penalty0
  2886--2896, 2016.

\bibitem[Lotspeich and Garcia(2025)]{LotspeichGarcia2025}
Sarah~C. Lotspeich and Tanya~P. Garcia.
\newblock Extrapolation before imputation reduces bias when imputing censored
  covariates.
\newblock \emph{Journal of Computational and Graphical Statistics}, 2025.
\newblock in press, doi: 10.1080/10618600.2024.2444323.

\bibitem[Bernhardt et~al.(2015)Bernhardt, Wang, and Zhang]{Bernhardtetal2015}
P.~W. Bernhardt, H.~J. Wang, and D.~Zhang.
\newblock Statistical methods for generalized linear model with covariates
  subject to detection limits.
\newblock \emph{Statistics in Biosciences}, 7\penalty0 (1):\penalty0 68--79,
  2015.

\bibitem[Royston(2007)]{Royston2007}
P.~Royston.
\newblock Multiple imputation of missing values: {F}urther update of ice, with
  an emphasis on interval censoring.
\newblock \emph{The Stata Journal}, 7\penalty0 (4):\penalty0 445--464, 2007.

\bibitem[Lotspeich et~al.(2022)Lotspeich, Grosser, and
  Garcia]{LotspeichGrosserGarcia2022}
Sarah~C. Lotspeich, Kyle~F. Grosser, and Tanya~P. Garcia.
\newblock Correcting conditional mean imputation for censored covariates and
  improving usability.
\newblock \emph{Biometrical Journal}, 64\penalty0 (5):\penalty0 858--862, 2022.

\bibitem[Meng(1994)]{Meng1994}
Xiao-Li Meng.
\newblock Multiple-imputation inferences with uncongenial sources of input.
\newblock \emph{Statistical Science}, 9\penalty0 (4):\penalty0 538–558, 1994.

\bibitem[Grembi(2022)]{riskCommunicator}
Jessica Grembi.
\newblock \emph{riskCommunicator: G-Computation to Estimate Interpretable
  Epidemiological Effects}, 2022.
\newblock URL \url{https://CRAN.R-project.org/package=riskCommunicator}.
\newblock R package version 1.0.1.

\end{thebibliography}


\section*{Supplementary Materials}

\begin{itemize}
    \item \textbf{Additional derivations, tables, and figures:} The supplemental figures and tables referenced in Sections 2--5 are available online at \url{https://github.com/sarahlotspeich/speedyCMI/blob/master/Supplementary_Materials.pdf} as Supplementary Materials.
    \item \textbf{R-package for imputation:} An R package \textit{speedyCMI} that implements the parametric conditional mean imputation methods described in this article is available at \url{https://github.com/sarahlotspeich/speedyCMI}.
    \item \textbf{R code and data for simulation studies and analysis:} The R scripts to replicate the simulation studies from Section~\ref{sec:sims} and  data analysis from Section~\ref{sec:framingham} are available at \url{https://github.com/sarahlotspeich/speedyCMI}.
\end{itemize}

\end{document}